# Wind environment analysis of ground-based optical observatory

**Taoran Li[1], Xiaojun Jiang[1,2]**

[1] Key Laboratory of Optical Astronomy, National Astronomical Observatories, Chinese Academy of Sciences, Beijing, 100101, China
[2] University of Chinese Academy of Sciences, Beijing,100049, China
E-mail: litaoran@bao.ac.cn

**Abstract**

The telescopes and the infrastructures may alter the local wind environment around the observatory and further affect the observing environment. After the completion of site testing, it is necessary to analyze the wind environment of the entire site and plan the telescope layout to make use of the excellent conditions scientifically and rationally. Taking a typical observatory as an example, the effect of topographical features on wind environment and the mutual interference between telescope enclosures are analyzed by using Computational Fluid Dynamics (CFD) method. The CFD simulations are compared with the seeing data from Differential Image Motion Monitor (DIMM), the results are in good agreement, which verifies the effectiveness of the CFD method. The results of wind environment analysis can provide reasonable suggestions for site layout and construction, improving the observing environment and the image quality.

## 1. Introduction

The observing environment of a ground-based optical observatory, one of the key factors affecting the image quality (Swapan 2007), consists of the natural environment (e.g., weather conditions, astronomical seeing) and local surroundings of the enclosure (e.g., temperature, wind speed and direction, turbulence).

Site testing is an essential work before the construction of a ground-based observatory. To search for a site with excellent natural environment requires long-term and arduous monitoring and research in multi-field such as astronomy, meteorology and geography. Usually, more than one telescope is built to take advantage of an excellent site, and other infrastructures are arranged, such as the Mauna Kea Observatories in Hawaii and the Observatorio del Roque de los Muchachos in La Palma, which has more than ten telescopes and auxiliary facilities. In addition, several telescopes will be installed at two new sites in west China: Lenghu (Deng et al. 2021) and Muztagh-ata (Xu et al. 2020).

However, architecture inevitably alters the microclimate in its vicinity (Blocken & Jan 2004). Interference effects may occur between several telescopes and facilities (Michael 1994), resulting in increased thermal equilibrium time and turbulence in the flow around the enclosure. On the other hand, the local wind field variations caused by topographic relief will make the observing environments around each telescope different. Therefore, wind environment directly affects the observing environment of the ground-based optical observatory, and is closely related to the image quality.

This paper focuses on the wind environment analysis of a typical ground-based optical observatory to fully exploit the potential of the excellent seeing conditions of the selected site and provide advice on observatory planning. Using Computational Fluid Dynamics (CFD) analyses based on enclosures and terrains, different factors affecting the wind environment are studied. This paper is organized as follows: Section 2 describes the influences of the irregular terrain and the surface roughness on wind environment. The interference effect between two enclosures is presented in Section 3. Section 4 compares the CFD results and the seeing data of two telescopes to verify the simulation method. A summarize is presented in the last section.

## 2. Topographical features influence on wind environment

The influences of telescopes layout and local surroundings are mainly considered in wind environment analysis. In this paper, the wind environment refers to the local wind field around the observatory. It is affected by the prevailing wind speed and direction, as well as the terrain and surface roughness.

Irregular terrain can make a chaotic wind environment. The influence of its fluctuations is more important than the surface roughness, which changes the wind speed and direction, and affects the stability of the wind field, increasing the possibility of turbulence (Abdel-Gawad & Zoklot 2006). Optical observatories are usually located in mountainous regions with higher altitudes, where the terrain is undulating and the atmospheric flow needs to overcome considerable ground friction, causing distortion accordingly. Surface roughness indicates the influence of the ground surface on atmospheric movement, which is one of the important factors that change the near-surface boundary layer. The influences of the above geographical features on the wind environment are described in this section.

The shape of the enclosure is negligible due to the large fluid domain for wind environment analysis. It usually affects the turbulence inside the dome. Therefore, in this paper, the most common type of enclosure is selected, i.e., hyper-hemisphere of the upper part and cylinder of the lower part.

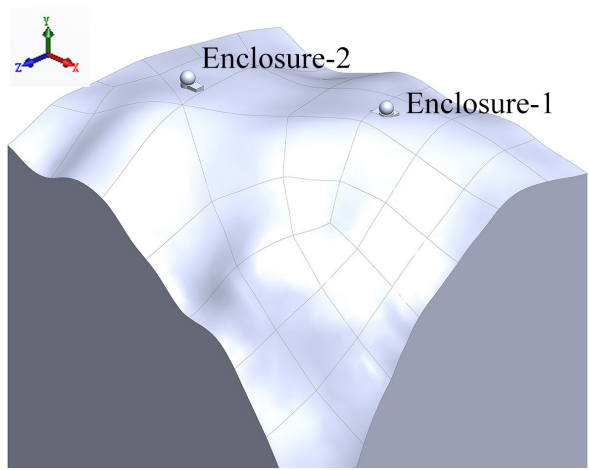

Fig. 1 3-D model of a typical observatory with two enclosures

Figure 1 shows the three-dimensional (3-D) model of an astronomical observatory with undulated terrain and two enclosures. The original file of a 620m×700m Digital Elevation Map (DEM) at 10m resolution was modified to be imported into the CFD software. Unsteady pressure-based Navier-Stokes simulations were performed using the k-epsilon turbulence model. An incoming velocity of 5m/s is specified directly to the windward slope with a 5% turbulence intensity.

The turbulence intensity and the vorticity are two common physical quantities in fluid mechanics, describing the relative intensity of fluctuating wind and the local spinning motion of flow fluid, respectively. The turbulence intensity is defined as the ratio of the Root Mean Square (RMS) of the turbulent velocity fluctuations to the mean velocity and the vorticity is the curl of the velocity vector[①]。

### 2.1 Terrain influence

Terrain undulations will influence the near-surface wind field and change the observing environment around the enclosure. Generally, the air is squeezed and accelerated by the ground when passing the windward slope, maximizing the wind speed at the summit and causing severe turbulence (Bowen & Lindley 1977). The telescopes located at the summit are influenced by this acceleration effect, increasing the wind speed and turbulence intensity, which could be detrimental to the observing environment. Several analyses are performed with a typical ground-based optical observatory to account for the impact of the local topography, including terrain slope influence, wind environment on the leeward mountainside and terrain influence on telescope layout.

### 2.1.1 Terrain slope

By analyzing the influence of terrain slope, combined with the dominant wind direction, a proper position for telescope could be given to avoid excessive turbulence and wind speed on the windward slope.

Three cases of 2-D simulation were used in wind environment analysis of slope influence, including steep slope, gentle slope and flat land, with the inclination angles of 47°, 25° and 0°, respectively. Figure 2 depicts the topographies of three cases and the distribution of turbulence intensity and wind speed. The topography of steep slope case is a cross-section through Enclosure1 of the 3-D model (in Figure 1), the other two cases (Figure 2b and 2c) are transformed from the topography in Figure 2a It could be noticed that the slope angle of the windward is linearly correlated to the wind speed, as well as the turbulence

---

[①] ANSYS, Inc. (2018), "ANSYS Fluent User's Guide, Release 19.0".

intensity, which verifies the acceleration effect mentioned above. The image quality may be affected on the leeward side, especially when the telescope's elevation angle is below 50°.

To avoid the "summit effect", the wind environment on the leeward mountainside was analyzed with the same topography in Figure 2a. The telescope was moved along the windward direction, about 150m away from the top. Figure 3 shows the distribution of turbulence intensity and wind speed. Compared with Figure 2a, the turbulent range on the leeward side of the enclosure is much smaller. The maximum affected elevation angle of telescope is only 20°.

Three points on windward, top and leeward of the enclosure were set for quantitative analysis. Results of steep

### 2.1.2 Leeward mountainside

slope case and leeward mountainside analysis are both given in table 1. It shows that the wind speed at the windward side decreases from 10.71m/s to 2.85m/s because of eliminating the "summit effect", while the turbulence intensity increases. In terms of turbulence intensity, this result could prove the opinion in Bely's book (Bely 2003) that the telescope should be placed as close to the ridge to sit in the unperturbed flow.

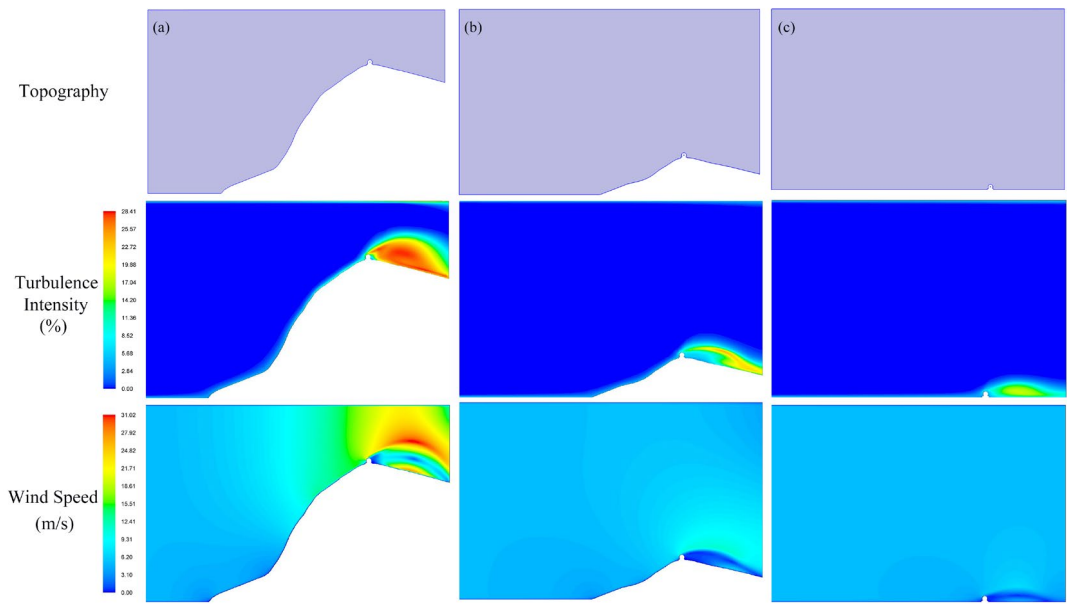

Fig. 2 Topography of slope influence analysis models and contours of turbulence intensity and wind speed, (a) steep slope; (b) gentle slope and (3) flat land.

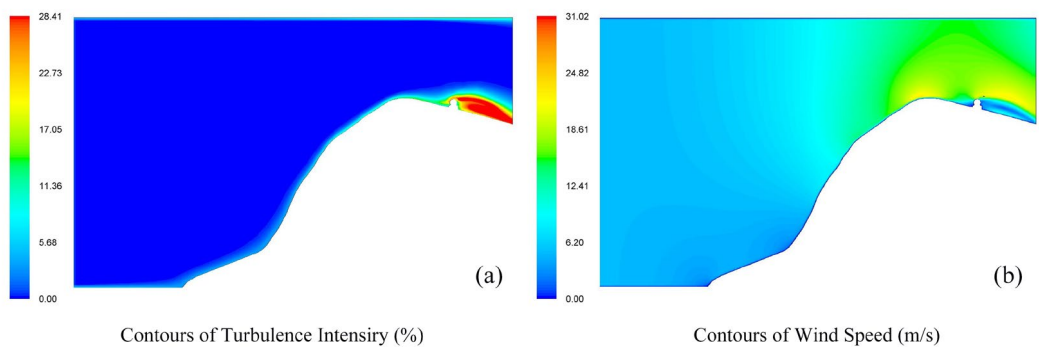

(a) Contours of Turbulence Intensiry (%)   (b) Contours of Wind Speed (m/s)

Fig. 3 Wind environment analysis of the telescope on the leeward mountainside. Contours of turbulence intensity (a) and Wind speed (b).

Table. 1 Wind fields at different sampling points of the cases of steep slopes and leeward mountainside

| Location of enclosure | Wind speed (m/s) | Turbulence intensity (%) |
| --- | --- | --- |

|  | Point1 | Point 2 | Point 3 | Point 1 | Point 2 | Point 3 |
|---|---|---|---|---|---|---|
| Leeward mountainside | 2.85 | 19.67 | 3.36 | 17.13 | 16.31 | 17.00 |
| Summit of steep slope | 10.71 | 19.68 | 0.40 | 9.34 | 15.70 | 9.63 |

In summary, both the accelerated effect and the unperturbed flow should be considered, combined with the specific goals to select the proper location of the telescope.

### 2.1.3 Terrain influence on telescope layout

A 3-D wind environment model based on Figure 1 was built to study the terrain influence on telescope layout (Figure 4a). The inlet velocity is in -z direction (blue arrow in Figure 4b). The outlet, top and two sides of the fluid domain are treated as the boundary condition of outflow, no-slip wall and symmetry, respectively. The numerical input parameters are the same as described in section 2.1.1.

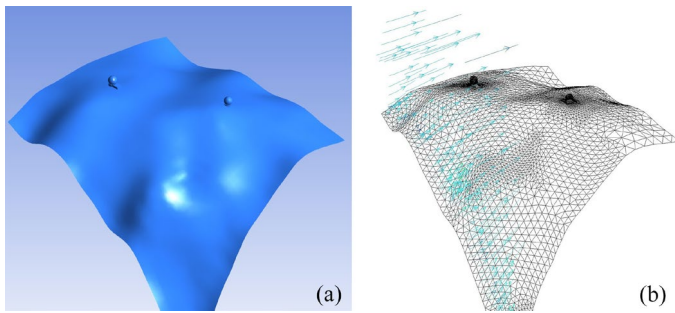

Fig. 4 3-D wind environment model, terrain and telescope geometry (a) and wind direction diagram (b).

The surface turbulence intensity distribution is shown in Figure 5, indicating the impact of terrain on wind environment. High turbulent areas can be found at the leeward side of two telescopes and the ground surface with the steepest slope. In addition, due to the complex topographical in the windward zone (+z direction) of Enclosure1, the turbulence intensity is about 8%~10%, which is significantly higher than other locations. As described in Figure 4b, the higher grid density also illustrates the undulated topography here. On both sides of Enclosure1 (Areas A and B in Figure 5), the grid is sparse and the terrain is relatively flat. Therefore, locations A and B with lower turbulence intensity are more suitable for building a telescope, as well as the location of Enclosure2.

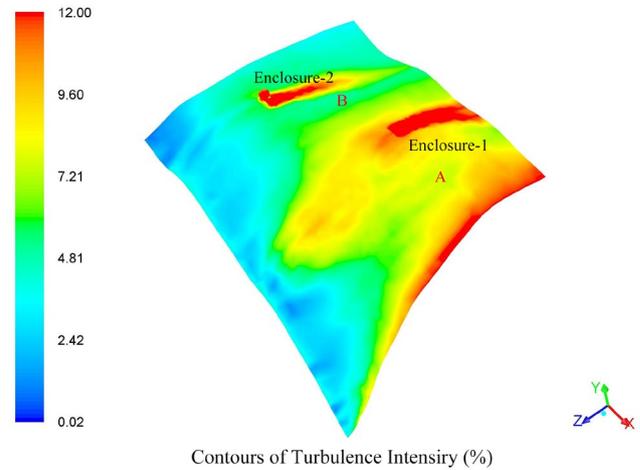

Fig. 5 Distribution of turbulence intensity on the ground surface

### 2.2 Surface roughness influence

The surface roughness reflects the influence on air movement. The wind speed near-surface is reduced due to the kinetic energy loss caused by topography and landform, such as sandy land or vegetation. It is inversely proportional to the height above the ground.

Table. 2 Roughness classification by Davenport and Wieringa

| Roughness classification | Landform description | Roughness length $l_0$ (m) |
|---|---|---|
| 1 | Open sea or lake, with a free fetch of several kilometers | 0.0002 |
| 2 | Mud flats, snow, no vegetation, no obstacles | 0.005 |
| 3 | Open flat terrain, grass, few isolated obstacles | 0.03 |
| 4 | Low crops, occasional obstacles | 0.10 |
| 5 | High crops, scattered obstacles | 0.25 |
| 6 | Parkland, bushes, numerous obstacles | 0.5 |
| 7 | Regular large obstacle coverage (suburb, forest) | 1.0 |
| 8 | City center with high- and low- rise buildings | ≥2.0 |

Surface roughness is usually expressed by roughness length $l_0$ (in meters) (Manwell et al. 2006). It is difficult to calculate the precise values for surface roughness. At present, many studies have given typical values of inhomogeneous terrain features. Table 2 shows the terrain classification by Davenport (Davenport et al. 2000) and Wieringa (Wieringa 1992) in terms of roughness length.

To study the surface roughness influence on wind environment, four landforms, including no vegetation, grass, low crops, and parkland, were selected to represent the various land cover of the typical optical observatory. The CFD software needs the roughness height $K_s$ as an input parameter, defined as,

$$K_s = 9.793 \times l_0 / C_s \text{ (Lee et al. 2018)} \quad (1)$$

where $C_s$ is the roughness constant with the default value of 0.5. Four landforms and roughness height in use are given in Table 3.

Table. 3 Landforms and roughness height values

| Landform of the site | Roughness constant | Roughness length $l_0$ (m) | Roughness height $K_s$ (m) |
|---|---|---|---|
| No vegetation | 0.5 | 0.005 | 0.098 |
| Grass | 0.5 | 0.030 | 0.588 |
| Low crops | 0.5 | 0.100 | 1.959 |
| Parkland | 0.5 | 0.500 | 9.793 |

In the gentle slope case of section 2.1.1, ten sampling points on two sides of the enclosure were set at the height of ten meters from the ground. The distance between each point is 10 meters, as shown in Figure 6.

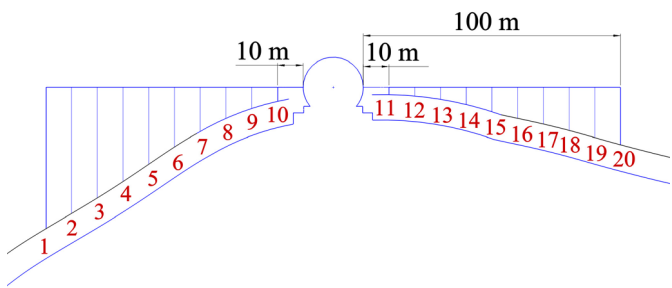

Fig. 6 Diagram of the sampling points

Figures 7 and 8 are wind speed and turbulence intensity curves, respectively. The horizontal axis is the distance from the enclosure, and the negative value represents the sampling points on the windward side of the enclosure. In this area, the variations of wind speed and turbulence intensity curves of different landforms show a good correlation with the surface roughness. A rougher surface causes a lower wind speed and a higher turbulence intensity. In the range of 10m~30m of the leeward side, the surface roughness has a minimal influence. As the distance increases, the curves become complicated. The variations are dominated by enclosure and undulated terrain.

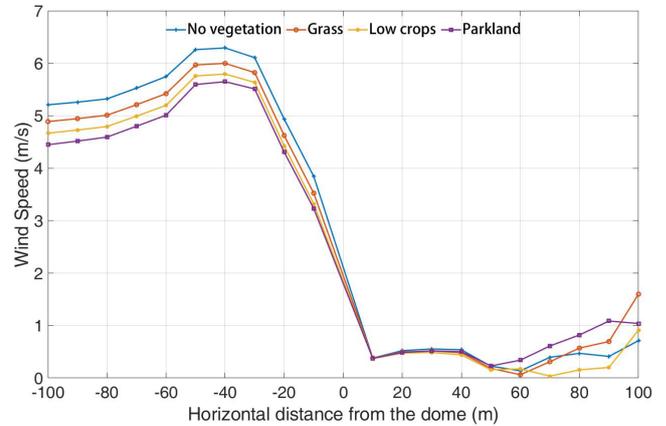

Fig. 7 Wind speed curves of four landforms

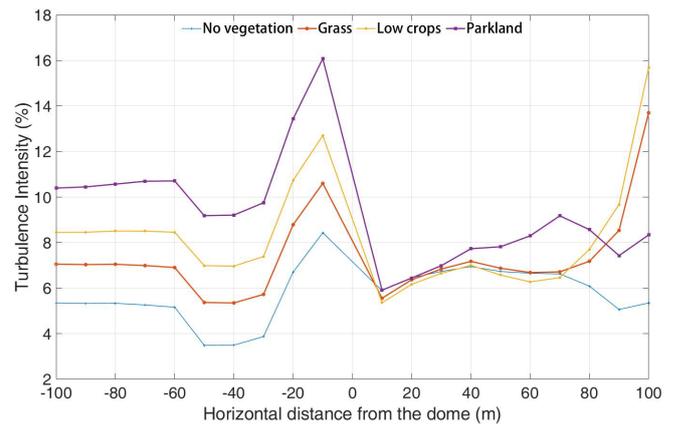

Fig. 8 Turbulence intensity curves of four landforms

The surface roughness mainly affects the wind environment of the windward side. A smoother terrain surface will make a better observing environment. The landforms of all observatories with excellent observing conditions are no vegetation or bare lands, such as the Saishiteng Mountain near Lenghu in China, Mauna Kea in Hawaii (Morrison et al. 1973) and La Silla in Chile. In addition, the surface roughness of the ice surface is extremely low. Dome A in Antarctic is considered to be the best site on Earth (Ma et al. 2020).

## 3. Observatory layout influence on wind environment

For a ground-based observatory, interference effects may occur between the enclosures. The windward surface of an enclosure will obstruct the airflow, inevitably changing the microclimate in its vicinity. The interaction between wind fields and adjacent buildings needs to be considered.

*3.1 Flow around a circular cylinder for enclosure*

The flow around a circular cylinder means that when the wind passes through an obstacle, a strong vortex will appear at the leeward side and gradually weaken along the downwind direction. Under certain conditions, the Karman vortex street (Karman 2013) will be generated, forming alternating vortices. The primary impact on astronomical observation is that the disturbance of the wake flow will cause wind environment variation around the telescopes in wake area (Figure 9), deteriorating dome seeing.

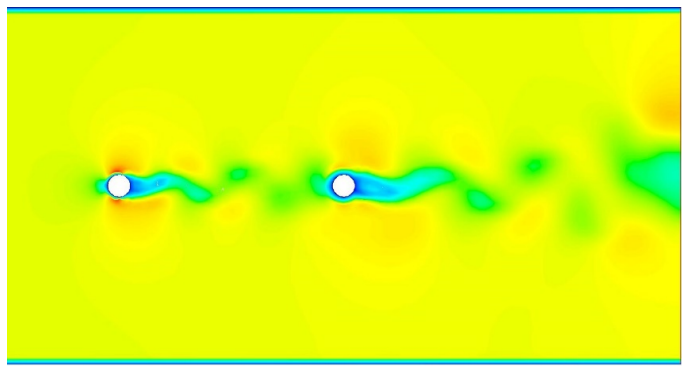

Fig. 9 Simulation of flow around circular cylinders with two enclosures

The wake intensity is weakening with distance from the front enclosure increase. The length of the wake region is usually ten times the height of the obstacle (CIMO 2010). Therefore, along the prevailing wind direction, the distance between the two enclosures should be larger than ten times the height of the previous enclosure.

*3.2 Interaction between two enclosures*

The 3-D CFD model in Section 2.1.3 was used to analyze the interaction between two enclosures, thereby providing suggestions on telescope layout. Enclosure1 and Enclosure2 represent the two telescopes (shown in Figure 10), with the height of 28 meters and 27 meters, respectively. The distance and the altitude difference between them are about 290 meters and 7 meters, respectively. The interaction analysis is performed with a wind speed of 5m/s and the wind direction along the two enclosures.

Figure 11 shows the wind speed distribution and flow pathlines in the cross-section of two enclosures. The wind speed on the windward side of Enclosure1 is reduced to 3.5m/s, due to the shelter provided by Enclosure2. The lower wind speed is conducive to wind load on a telescope, but the efficiency of natural ventilation will decrease. The pathline in Figure 11b shows that the airflow on the windward side of Enclosure1 is relatively smooth and steady, the impact of Enclosure2 may not be obviously seen.

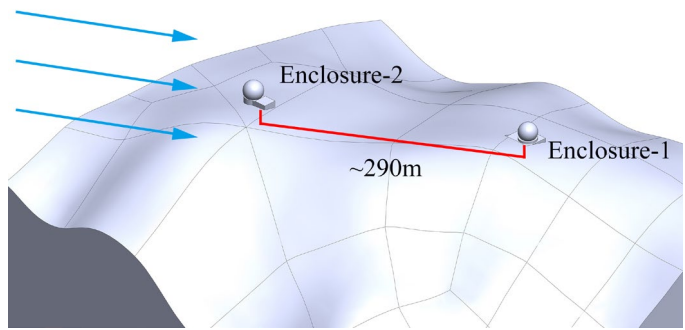

Fig. 10 The analysis model of interaction between two enclosures. The distance between the two enclosures is about 290 meters. The wind direction is represented in arrows. Enclosure1 has a height of 27 meters and a diameter of 23 meters. Enclosure2 has a height of 28 meters and a diameter of 25 meters.

The distributions of turbulence intensity and vorticity magnitude are shown in Figure 12 and Figure 13, respectively, to analyze the impact of turbulence generated by Enclosure2 on Enclosure1. The scatter plot data are displayed as a collection of points in the cross-section of two enclosures. The horizontal axis is the position value from the leftmost in Figure 12a. Both the turbulence intensity and the vorticity start increasing rapidly at 140 meters (the leeward side of Enclosure2). As the distance gets farther, the values gradually decrease until they increase again at 400 meters (the windward side of Enclosure1).

As shown in Figure 14, the sampling rectangles are centered on top of each enclosure, with a length of 3D ("D" represents the diameter of the dome) and width of 2D. The intervals of sampling points are 0.3D and 0.2D, respectively. Then, the available points are 87 and 89 in total for Enclosure2 and Enclosure1 because of terrain and enclosure overshadowed. Table 4 gives the measured data of sampling points. The average value of turbulence intensity around Enclosure1 is about 2.15 times that of Enclosure2, and the RMS of the difference between them is 3.51. If the windward side is considered alone, the ratio of the average value of turbulence intensity becomes 5.36. Compared with Enclosure2, the turbulence intensity on the windward side of Enclosure1 is much higher and the difference of vorticity magnitude is small, indicating that the wind environment near Enclosure1 is mainly affected by the fluctuating wind caused by Enclosure2.

Therefore, the telescopes in wake area may still be affected by the windward enclosure even the distance between them is larger than ten times the height of the previous one. Usually, the strategies of layout optimization are as follows,

(1) If the angle between the line of two enclosures and the prevailing wind is less than $\arctan\left(\dfrac{R_1 + R_2}{L}\right)$, where L is the distance between two enclosures, R1 and R2 are the diameters, it is necessary to analyze the observing environment based on terrain to determine the interference.

(2) A convenient plan of rearrangement is to move the telescope along the direction perpendicular to the prevailing wind, away from the wake area of the windward enclosure.

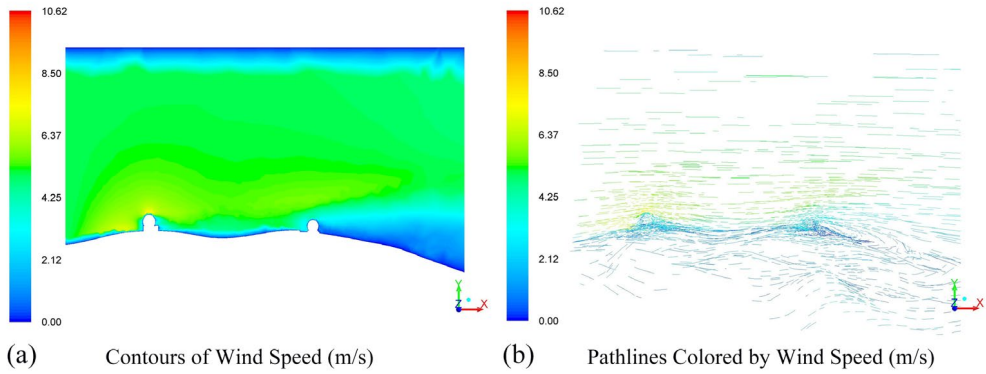

(a)  Contours of Wind Speed (m/s)  (b)  Pathlines Colored by Wind Speed (m/s)

Fig. 11 Wind speed distribution in the cross-section of two enclosures, (a) contours of wind speed, (b) flow pathlines. The wind direction is parallel with +z axis (from left to right).

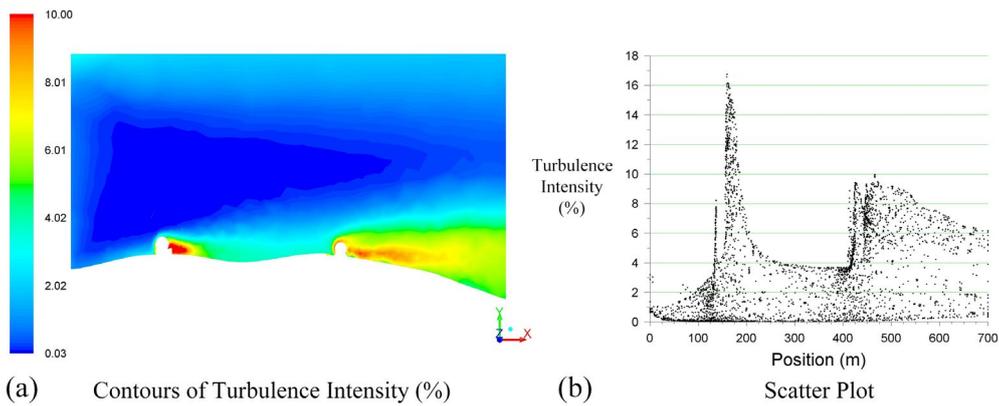

(a)  Contours of Turbulence Intensity (%)  (b)  Scatter Plot

Fig. 12 Turbulence intensity distribution in the cross-section of two enclosures, (a) contours of turbulence intensity, (b) scatter plot. The wind direction is parallel with +z axis (from left to right).

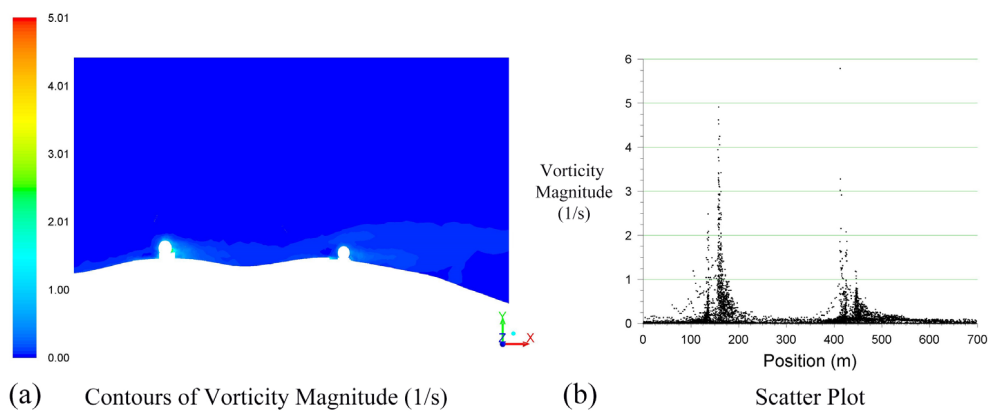

(a)  Contours of Vorticity Magnitude (1/s)  (b)  Scatter Plot

Fig. 13 Vorticity magnitude distribution in the cross-section of two enclosures, (a) contours of vorticity magnitude, (b) scatter plot. The wind direction is parallel with +z axis (from left to right).

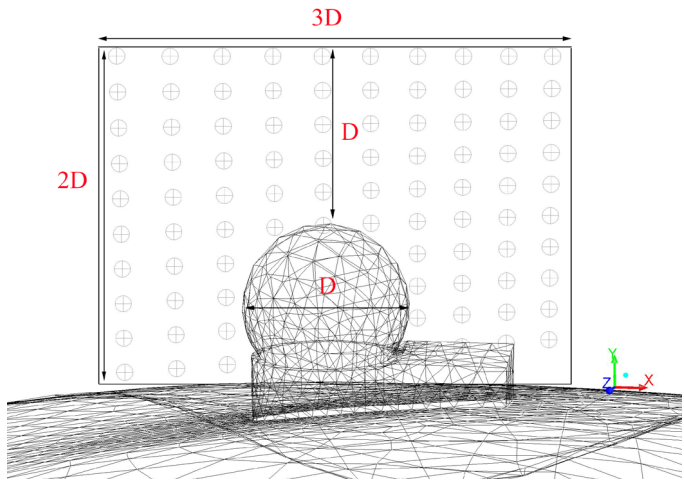

Fig. 14 Diagram of sampling points. The sampling rectangles are centered on the top of each enclosure, with a length of 3D ("D" represents the diameter of the dome) and 2D. The sampling intervals are 0.3D and 0.2D, respectively.

Table. 4 Comparison of turbulence intensity and vorticity magnitude around two enclosures

| Parameters | Turbulence intensity (%) | Vorticity magnitude (1/s) |
|---|---|---|
| RMS error (Enclosure1-Enclosure2) | 3.51 | 0.26 |
| Average_Enclosure1 / Average_Enclosure2 | 2.15 | 0.77 |
| Average_Enclosure1 / Average_Enclosure2 (Windward area) | 5.36 | 1.10 |

## 4. Verification of simulation method

The Telescopio Nazionale Galileo (TNG) and William Herschel Telescope (WHT) located at ORM in Spain have conducted measurements of observing environment for many years and published the comparison results of Differential Image Motion Monitor (DIMM) (Munoz-Tunon et al. 1997 and Molinari et al. 2012). In this section, the wind environment of TNG and WHT was analyzed to verify the validity of the CFD method.

Figure 15 shows the locations of two DIMMs. The TNG-DIMM is located about 108 meters southwest of TNG and 15 meters above the ground. The WHT-DIMM is located about 50 meters northeast of WHT and its altitude difference with TNG-DIMM is nearly 50 meters. Two cubes with the center of each DIMM and a side length of 9 meters were drawn to collect the turbulence intensity. Both horizontal and vertical intervals of the sampling points are 3 meters, given 64 points in total.

The median results of simulated turbulence intensity under different wind directions are summarized in Table 5. The statistical distribution of wind direction was derived from the average of TNG weather station over seven years (1998-2004) (Lombardi et al. 2007). Since the wind direction significantly influences on turbulence intensity, it is unreasonable to compare the results under different wind directions directly. The wind direction weighted needs to be considered. The weighted turbulence intensity, conveniently calculated by rounded results of wind direction percentage (WDP), is defined as,

$$\text{Weighted turbulence intensity} = \text{median}\{WDP_N*(TI_N)\bigcup...\bigcup WDP_{NW}*(TI_{NW})\} \quad (2)$$

where TI is the set of turbulence intensity for eight wind directions, the symbol $\bigcup$ represents the union operator. Table 6 shows the comparison of DIMM and the simulated results, with WHT-DIMM values slightly higher. The ratio of turbulence intensity at TNG-DIMM and WHT-DIMM is 0.94, similar to the results of DIMM values (0.91~0.93), indicating the good agreement of the two independent measures and proving the validity and accuracy of the CFD method.

Table. 5 Simulated turbulence intensity of TNG-DIMM and WHT-DIMM in different wind direction

| Wind direction | N | NE | E | SE | S | SW | W | NW |
|---|---|---|---|---|---|---|---|---|
| Wind direction percentage (%, Rounded) | 6 | 20 | 13 | 10 | 14 | 13 | 11 | 13 |
| Median turbulence intensity at TNG-DIMM (%) | 2.65 | 5.92 | 10.73 | 4.12 | 3.62 | 2.28 | 1.92 | 2.12 |
| Median turbulence intensity at WHT-DIMM (%) | 3.14 | 3.36 | 3.87 | 5.43 | 6.74 | 7.22 | 2.15 | 2.36 |

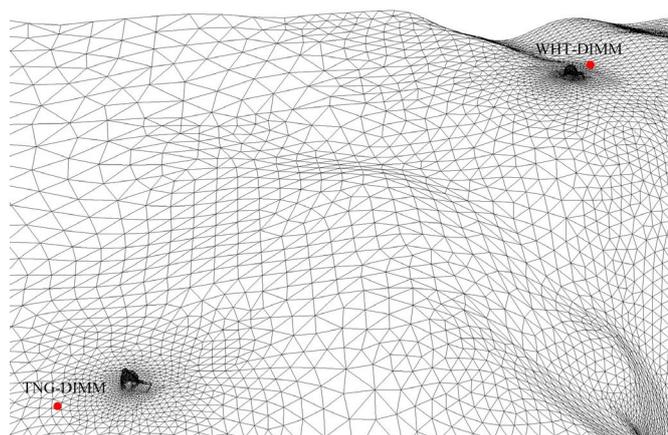

Fig. 15 Diagram of the locations of two DIMMs. WHT-DIMM is 1100 m northeast far from TNG-DIMM.

Table. 6 Comparison of DIMM and simulated turbulence intensity

| Median results | TNG /" | WHT /" | Ratio |
|---|---|---|---|
| DIMM (1994~1995) (Munoz-Tunon et al. 1997) | 0.64 | 0.69 | 0.93 |
| DIMM (2011~2012) (Molinari et al. 2012) | 0.86 | 0.94 | 0.91 |
| DIMM (2020~2021)② | 0.69 | 0.74 | 0.93 |
| Turbulence intensity in CFD (Wind direction weighted) | 3.62 | 3.84 | 0.94 |

## 5. Conclusion

In this paper, CFD simulations are presented to analyze the wind environment of the ground-based optical observatory. The influences of topographical features on wind environment are analyzed based on the typical terrain and landforms. The wind speed is accelerated by terrain as the slope increases, bringing turbulent airflow at the summit. The accelerated effect and the unperturbed flow should be considered to select the proper location of one telescope, avoiding the turbulence caused by natural terrain. A study on the influence of surface roughness reveals the impact of landforms on wind speed and turbulence intensity near-surface. Some observatories have planted many trees, leading to the increase of water vapor, as well as the high turbulent airflow. That is going against the principle of site selection. For the interaction problems by adjacent buildings, the comparisons of wind speed and turbulence intensity around two enclosures are performed. Due to the inevitable interference by the windward enclosure, it is reasonable for every telescope project to analyze the observing environment based on terrain. Moving the telescope away from the wake area is a convenient plan of rearrangement. A comparison of the wind environment analysis with TNG and WHT helps validate the CFD methods used for this work. The ratio of weighted turbulence intensity near TNG and WHT (0.94) is close to the DIMM values (0.91~0.93). The developed method of wind environment analysis will provide references for the rational planning of an optical observatory to take advantage of the excellent seeing conditions and improve the image quality.

This work is supported by the National Natural Science Foundation of China (U1831209).

We would like to thank the anonymous referee for the hard work and valuable suggestions that helped improve the manuscript.

## References

[1] Swapan K.S. 2007, Diffraction-Limited Imaging with Large and Moderate Telescopes. (Singapore: World Scientific Publishing Co.Pte.Ltd)
[2] Deng, L.C., Yang F., Chen, X.C., et al. 2021, *Nature*, 596, 353
[3] Xu, J., Esamdin, A., Hao,J.X., et al. 2020, *Res. Astron. Astrophys*, 20, 086
[4] Blocken, B., & Jan C. 2004, *Journal of Thermal Envelope and Building Science*, 28, 2
[5] Michael,W. S. 1994, *Proc. SPIE* 2199, 465
[6] Abdel-Gawad, A.F., & Zoklot, A.S.A. 2006, *WIT Transactions on Ecology and the Environment*, 93, 433
[7] Bowen, A.J., & Lindley, D. A. 1977, *Boundary-Layer Meteorol*, 12, 259
[8] Bely, P.Y., 2003, The Design and Construction of Large Optical Telescopes. (New York: Springer)
[9] Manwell, J.F., McGowan, J.G., & A.L. Rogers. 2006, Wind Energy Explained: Theory, Design and Application. (West Sussex: John Wiley & Sons)
[10] Davenport, A.G., Grimmond, C.S.B., Oke, T.R., et al. 2000, *Preprints of the Twelfth American Meteorological Society Conference on Applied Climatology*, 4B, 96
[11] Wieringa, J. 1992, *Wind Engineering*, 41, 1
[12] Lee, Y.C., Chang, T.J., & Hsieh, C.I. 2018, *Sustainability*, 10, 1190
[13] Morrison, D., Murphy, R.E., Cruikshank, D.P. 1973, *PASP*, 85, 255
[14] Ma, B., Shang, Z.H., Hu, Y., et al. 2020, *Nature*, 583, 7818
[15] Karman, T.V. 2013, *Progress in Aerospace Sciences*, 59,13.
[16] Commission for Instruments and Methods of Observation (CIMO). 2010, A*bridged final report with resolutions and recommendations* (Helsinki: World meteorological organization), No.1064
[17] Munoz-Tunon, C., Vernin, J., & Varela, A. M. 1997, *A&AS*, 125,183
[18] Molinari, E., de Gurtubai, A. G., della Valle, A., et al. 2012, *Proc. SPIE* 8444, 844462
[19] Lombardi, G., Zitelli, V., Ortolani, S., et al. 2007, *PASP*, 119, 292

---

② Based on ORM DIMM Seeing website by Nordic Optical Telescope. http://www.not.iac.es/weather/index.php